# Phase change materials for nano-polaritonics: a case study of hBN/VO$_2$ heterostructures

S. Dai[1,2,†,*], J. Zhang[3,†], Q. Ma[4], S. Kittiwatanakul[5], A. S. McLeod[6], X. Chen[3], S. N. Gilbert Corder[3], K. Watanabe[7], T. Taniguchi[7], J. Lu[5], Q. Dai[8], P. Jarillo-Herrero[4], M. K. Liu[3]*, D. N. Basov[6]*.

[1]*Department of Physics, University of California, San Diego, La Jolla, California 92093, USA*

[2]*Department of Electrical & Computer Engineering, The University of Texas at Austin, Texas 78712, USA*

[3]*Department of Physics, Stony Brook University, Stony Brook, New York 11794, USA*

[4]*Department of Physics, Massachusetts Institute of Technology, Cambridge, Massachusetts 02215, USA*

[5]*Department of Physics, University of Virginia, Charlottesville, Virginia 22904, USA*

[6]*Department of Physics, Columbia University, New York, New York 10027, USA*

[7]*National Institute for Materials Science, Namiki 1-1, Tsukuba, Ibaraki 305-0044, Japan*

[8]*Division of Nanophotonics, CAS Center for Excellence in Nanoscience, National Center for Nanoscience and Technology, Beijing 100190, People's Republic of China*

*Correspondence to: sdai@utexas.edu, mengkun.liu@stonybrook.edu, db3056@columbia.edu

[†]These authors contributed equally to this work

Abstract:

Polaritonic excitation and control in van der Waals (vdW) materials exhibit superior merits than conventional materials and thus hold new promise for exploring light matter interactions. In this work, we created vdW heterostructures combining hexagonal boron nitride (hBN) and a representative phase change material – vanadium dioxide ($VO_2$). Using infrared nano-spectroscopy and nano-imaging, we demonstrated the dynamic tunability of hyperbolic phonon polaritons in hBN/VO2 heterostructures by temperature control in a precise and reversible fashion. The dynamic tuning of the polaritons stems from the change of local dielectric properties of the $VO_2$ sublayer through insulator to metal transition by the temperature control. The high susceptibility of polaritons to electronic phase transitions opens possibilities for applications of vdW materials in combination with correlated phase change materials.

Van der Waals (vdW) heterostructures[1, 2] assembled together by vdW forces offer a versatile novel platform to control physical properties and quantum phenomena[3]. Specifically, as building blocks for the heterostructures, vdW materials have shown innovative electronic, optical and magnetic properties. They harbor Dirac[4] and valley[5] electrons, support superconductivity[6, 7] and ferromagnetism[8, 9], emit light[10, 11] and exhibit topologically protected states[12, 13], as well as many other effects. Stacking various vdW materials into heterostructures can intertwine these properties in a hybrid and controllable fashion for advanced functionalities[1, 2]. In addition, phase change behaviors as important basics for energy storage[14, 15], reconfigurable electronics and optics[16], have just been explored in vdW systems recently[17, 18]. Therefore, in addition to some preliminary approaches[19-21], the exploration of heterostructures composed of vdW crystals and phase change materials is imperative.

In this work, we build a novel vdW heterostructure (Figure 1a) by combining hexagonal boron nitride (hBN) and a canonical phase change material vanadium dioxide ($VO_2$) for the control of polaritons[22, 23]. Polaritons are hybrid light-matter modes involving collective oscillations of electromagnetic dipoles in materials. One representative polaritonic vdW crystal is hBN where hyperbolic phonon polaritons ($HP^2$s) are supported[24-31]. $HP^2$s in hBN originate from anisotropic phonon resonances which naturally facilitate Type I ($\varepsilon_z < 0$, $\varepsilon_t > 0$) and Type II ($\varepsilon_z > 0$, $\varepsilon_t < 0$) hyperbolicity. This natural hyperbolicity offer advantages over previous efforts using metamaterials[32, 33] due to the high momentum cutoff of light set by the interatomic spacing in natural lattices[34, 35]. Therefore, the hyperbolicity in hBN holds promising applications in subdiffractional focusing[34, 35], emission engineering[36] and light steering[26, 30, 33]. Despite the above merits, $HP^2$s in hBN are insensitive to external stimuli due to the insulating nature of the crystals. Dynamic tuning of $HP^2$s in hBN remains limited to electrostatic gating in graphene/hBN heterostructures[37] based primarily on the electronic and plasmonic properties of graphene. Here we report on nano-spectroscopy and nano-imaging results revealing the dynamic tuning of phonon resonances and HP$^2$s in hBN/VO$_2$ heterostructures via the control of temperature. We remark that optical switching of surface phonon polaritons with phase change materials has been previously achieved in Ge$_3$Sb$_2$Te$_6$ on a quartz substrate[38]. However, dynamic tuning of polaritons and especially the hyperbolicity haven't been reported therein. We hereby demonstrate the hybridization between polaritons in hBN and insulator-to-metal transition (IMT) in VO$_2$ via nano-imaging the heterostructures inside the Type II hyperbolic region. We also show the dynamic tuning of spectroscopic phonon resonances in both Type I and Type II region by varying the temperature of the heterostructure.

Scattering-type scanning near-field optical microscopy (s-SNOM) is used to study the hybrid polariton and phase transition in hBN/VO$_2$ heterostructures. Briefly, we illuminate the tip of an atomic force microscope (AFM) with infrared (IR) quantum cascade lasers (QCL, solid red arrow, Figure 1a) and collect the back-scattered IR signal (dashed red arrow). The scattered signal is demodulated at higher harmonics of the tip tapping frequency to yield essential near-field signal. The s-SNOM can record optical signals inaccessible by conventional optics with a resolution down to 10 nm (Methods) and has been extensively used to investigate nano-optical phenomena, including polaritons in vdW materials[22, 24, 39, 40] and nanoscale phase transitions[41, 42]. The VO$_2$ film studied in this work has a thickness of 50 nm and was deposited on a $[110]_R$ TiO$_2$ substrate to ensure a large conductivity jump (five orders of magnitude) during the IMT[41]. This VO$_2$/TiO$_2$ system also renders a unidirectional phase separation which permits an anisotropic electronic phase modulation[43].

The [110]$_R$ VO$_2$ undergoes anisotropic IMT over the temperature range from 320 K to 345 K[41]. The evolution of this IMT was first characterized with s-SNOM at the surface of VO$_2$ films: from the homogenous insulating state to metallic nucleation clusters, and finally unidirectional metallic stripes. These different stages of IMT with increasing temperature are plotted at the hBN/VO$_2$ interfaces in Figure 1a to Figure 1d. We note that the s-SNOM recorded IMT is nondispersive in the mid-IR as identical images were obtained regardless of the IR frequency (Figures 1d-f, hBN/VO$_2$ interfaces).

In the heterostructures of hBN on VO$_2$, nano-optical features of the phase transition were revealed at the hBN top surface via the hybridization between HP$^2$s and the IMT. Once illuminated with the IR light (Figure 1a, red arrow), the metalized AFM tip launches propagating HP$^2$s by bridging the momentum from photons to polaritons[22, 34, 35, 44]. The phase boundaries between the insulating and metallic state in hBN/VO$_2$ interfaces (Figures 1d-f) act as effective reflectors for HP$^2$s. Ring-shape polariton fringes appear on the hBN top surface (Figure 1b) due to standing wave interference between reflected and launched HP$^2$s. These fringes can connect with each other as the density of metallic nucleation cites increases at a higher temperature in the IMT of VO$_2$ (335 K, Figure 1c). As the temperature further increases, metallic stripes emerge on the surface of VO$_2$ and the corresponding HP$^2$s fringes align into a similar stripe shape on the hBN top surface (Figure 1d). Our interpretation of hybrid polariton-phase-transition for the observed results in Figures 1a-d is supported by the s-SNOM image at a shorter IR wavelength ($\lambda_{IR}$ = 6.4 µm, Figure 1e). At this IR frequency, the polariton fringes become narrower, revealing a shorter HP$^2$s wavelength, in accord with the frequency – momentum dispersion of HP$^2$s[24, 27]. The hybrid polariton-phase-transition is further verified in a control experiment: no evident nano-optical features can be observed at a much shorter IR wavelength ($\lambda_{IR}$ = 5.7 µm, Figure 1f) out-side-of the hyperbolicity regime. In comparison to previous works relying on topographic reflectors/launchers for nano-polaritons[24, 39, 40, 45], the conducting stripes in VO$_2$ here are predominantly electronic reflectors and do not reveal any evident topographic features in AFM measurements. The reflection of HP$^2$s can be attributed to local permittivity variations that are *electronically* formed at the phase boundaries in VO$_2$.

Having established the hybridization between polaritons in hBN and the IMT in VO$_2$, we now demonstrate the dynamic tuning of the phonon resonances by performing Fourier Transform Infrared Nano-spectroscopy (nano-FTIR) on the hBN/VO$_2$ heterostructure. The spectroscopic nano-FTIR response is averaged over the sample surface to avoid areal-dependent polaritonic effects (Figure 1). hBN supports *z* axis and *xy* plane phonon resonances in the mid-IR that are recorded as spectral peaks in the low (Figure 2a, Type I) and high (Figure 2b, Type II) frequency regions, respectively. At room temperature, the spectral features are similar to reported results in hBN/SiO$_2$ structures[37]. As the temperature increases, interesting spectral evolutions in the heterostructure of hBN/VO$_2$ can be observed in both Type I and Type II regions. In the Type I region (Figure 2a), the spectral intensity of the *z* axis phonon increases as VO$_2$ undergoes the IMT at the increasing temperature. In the Type II region (Figure 2b), on the contrary, the spectral intensity of the *xy* plane phonon decreases at the increasing temperature. While the intensity of the nano-FTIR spectra shows opposite trends in the two spectral regions, both the Type I and Type II resonances red-shift (black dotted arrows in Figures 2a and b) during IMT with increasing temperature. Therefore, both the intensity and the peak frequency of the hBN phonon resonance can be tuned by the control of temperature-dependent IMT in VO$_2$.

The tuning of phonon resonances stems from the modification of hyperbolic polaritons in hBN/VO$_2$ heterostructures by varying the temperature. We account for this effect by performing the simulation of frequency (ω) – momentum ($k_t$) dispersion (Figures 2c and d) in hBN/VO$_2$ heterostructures. Our nano-FTIR probes spectroscopic response of materials with in-plane momenta ($k_t$) that can be effectively coupled to the s-SNOM. The nano-FTIR spectra are therefore related to optical modes within a finite momentum region scoped by the time-averaged near-field coupling weight function[46] $G = <k_{xy}^2 e^{-2k_{xy}z_d}>_t$ (white dashed curves in Figures 2c, d). Here $z_d$ is the distance between sample surface and the simulated point dipole[46]. In hBN, multiple branches of HP$^2$s (shown as false color in Figures 2c, d) in Type I and Type II regions contribute to the $z$-axis and $xy$-plane phonon resonance observed in the nano-FTIR spectra, respectively (300 K, black curves in Figures 2a, b). The phase transition in VO$_2$ is modeled as a variable dielectric ground plane under the hBN crystal. The complex dielectric constants of VO$_2$ at the fully insulating and metallic phases are extracted from previous experiments[47]. As VO$_2$ undergoes the IMT, Type I HP$^2$s shift into the strong coupling region (white dashed curves) whereas in the Type II region, HP$^2$s shift outside of the region. This is evident by comparing the ω – $k_t$ dispersion of the hBN on insulating (false color in Figures 2c and 2d) and on metallic (red curves in Figures 2c and 2d) VO$_2$. Therefore, during the IMT, the coupling between the s-SNOM with more polaritonic modes can account for the increasing spectral intensity in the Type I region. In contrast, coupling with fewer polaritonic modes in the Type II region will cause the decrease of the spectral intensity. Moreover, with increasing metallicity, the hybrid HP$^2$s-IMT modes red-shift in both Type I and Type II region, in accordance with our experimental observations at the increasing temperature (Figures 2a-b).

The nano-imaging and nano-spectroscopy results augmented with the frequency – momentum dispersion simulations in Figures 1 and 2 demonstrate dynamic tuning of polaritons in heterostructures of hBN/VO$_2$ made possible by the phase transition. Effective redirection of HP$^2$s in hBN has also been achieved using metallic nucleation and stripes in VO$_2$ as an electronic reflector. The dynamic tuning of nano-polaritons and hyperbolicity in hBN can be attributed to the delicate control of the dielectric environment of the polaritonic medium using IMT in $VO_2$. Future studies will be directed towards ultrafast pump-probe control[48] of the tuning of nano-polaritons with phase change materials. We also envision the possibility of engineering local patterns[38] in phase change materials for memory polaritonics, polaritonic circuits, and other advanced nanophotonic functionalities[49]. Furthermore, the methodology we utilized in hBN/$VO_2$ can be extended for the tuning of nano-polaritons in heterostructures combing phase change materials with other vdW materials, including graphene[50], transition metal dichalcogenides[17], topological insulators[51] and black phosphorus[52] as well as metamaterials and metasurfaces where the polaritonic modes are also sensitive to the dielectric surroundings.

During the preparation of this manuscript, we became aware of another experimental work on phonon polaritons in hBN/VO2 heterostructures[53].

**Methods**

*Experimental setup*

The infrared (IR) nano-imaging and Fourier transform infrared nano-spectroscopy (nano-FTIR) experiments on hBN/$VO_2$ heterostructures were performed using a scattering-type scanning near-field optical microscope (s-SNOM). We use a commercial s-SNOM system (www.neaspec.com) based on a tapping-mode atomic force microscope (AFM) and a commercial AFM tip (tip radius ~ 10 nm) with a $PtIr_5$ coating. In the experiment, we illuminate the AFM tip by monochromatic quantum cascade lasers (QCLs) (www.daylightsolutions.com) and a broadband source from a difference frequency generation system (www.lasnix.com) to cover a frequency range of 900 – 2300 cm$^{-1}$. The s-SNOM nano-images were recorded by a pseudo-heterodyne interferometric detection module with an AFM tapping frequency 280 kHz and tapping amplitude around 70 nm. In order to remove the background signal, we demodulate the s-SNOM output signal at the 3$^{rd}$ harmonics of the tapping frequency.

*Sample fabrication*

The vanadium dioxide (VO$_2$) substrates studied in this report are deposited on [110]$_R$ TiO$_2$ substrates (1 cm × 1 cm) by temperature-optimized sputtering using the reactive bias ion beam deposition in a mixed Ar and O$_2$ atmosphere. These substrates show consistently anisotropic THz and DC conductivity[41]. A 1% ~ 2% tensile strain along c$_R$ axis is present in these substrates due to the lattice mismatch between TiO$_2$ and VO$_2$. Hexagonal boron nitride (hBN) crystals are mechanically exfoliated onto the VO$_2$/TiO$_2$ substrates.

**References:**

1. Geim, A.K. & Grigorieva, I.V. Van der Waals heterostructures. *Nature* **499**, 419 (2013).
2. Novoselov, K.S., Mishchenko, A., Carvalho, A. & Castro Neto, A.H. 2D materials and van der Waals heterostructures. *Science* **353** (2016).
3. Basov, D.N., Averitt, R.D. & Hsieh, D. Towards properties on demand in quantum materials. *Nature Materials* **16**, 1077 (2017).
4. Castro Neto, A.H., Guinea, F., Peres, N.M.R., Novoselov, K.S. & Geim, A.K. The electronic properties of graphene. *Reviews of Modern Physics* **81**, 109-162 (2009).
5. Xu, X., Yao, W., Xiao, D. & Heinz, T.F. Spin and pseudospins in layered transition metal dichalcogenides. *Nature Physics* **10**, 343 (2014).
6. Cao, Y. et al. Unconventional superconductivity in magic-angle graphene superlattices. *Nature* **556**, 43 (2018).
7. SUBRAMANIAN, M.A. et al. A New High-Temperature Superconductor: $Bi_2Sr_{3-x}Ca_xCu_2O_{8+y}$. *Science* **239**, 1015-1017 (1988).
8. Gong, C. et al. Discovery of intrinsic ferromagnetism in two-dimensional van der Waals crystals. *Nature* **546**, 265 (2017).
9. Huang, B. et al. Layer-dependent ferromagnetism in a van der Waals crystal down to the monolayer limit. *Nature* **546**, 270 (2017).
10. Mak, K.F., Lee, C., Hone, J., Shan, J. & Heinz, T.F. Atomically Thin ${\mathrm{MoS}}_{2}$: A New Direct-Gap Semiconductor. *Physical Review Letters* **105**, 136805 (2010).
11. Splendiani, A. et al. Emerging Photoluminescence in Monolayer MoS2. *Nano Letters* **10**, 1271-1275 (2010).
12. Hasan, M.Z. & Kane, C.L. Colloquium: Topological insulators. *Reviews of Modern Physics* **82**, 3045-3067 (2010).
13. Qi, X.-L. & Zhang, S.-C. Topological insulators and superconductors. *Reviews of Modern Physics* **83**, 1057-1110 (2011).
14. Sharma, A., Tyagi, V.V., Chen, C.R. & Buddhi, D. Review on thermal energy storage with phase change materials and applications. *Renewable and Sustainable Energy Reviews* **13**, 318-345 (2009).
15. Wuttig, M. & Yamada, N. Phase-change materials for rewriteable data storage. *Nature Materials* **6**, 824 (2007).
16. Wuttig, M., Bhaskaran, H. & Taubner, T. Phase-change materials for non-volatile photonic applications. *Nature Photonics* **11**, 465 (2017).
17. Wang, Y. et al. Structural phase transition in monolayer MoTe2 driven by electrostatic doping. *Nature* **550**, 487 (2017).
18. Cao, Y. et al. Correlated insulator behaviour at half-filling in magic-angle graphene superlattices. *Nature* **556**, 80 (2018).
19. Goldflam, M.D. et al. Tuning and Persistent Switching of Graphene Plasmons on a Ferroelectric Substrate. *Nano Letters* **15**, 4859-4864 (2015).
20. Kalikka, J., Zhou, X., Behera, J., Nannicini, G. & Simpson, R.E. Evolutionary design of interfacial phase change van der Waals heterostructures. *Nanoscale* **8**, 18212-18220 (2016).

21. Momand, J. et al. Atomic stacking and van-der-Waals bonding in GeTe–Sb2Te3 superlattices. *Journal of Materials Research* **31**, 3115-3124 (2016).
22. Basov, D.N., Fogler, M.M. & García de Abajo, F.J. Polaritons in van der Waals materials. *Science* **354** (2016).
23. Low, T. et al. Polaritons in layered two-dimensional materials. *Nature Materials* **16**, 182 (2016).
24. Dai, S. et al. Tunable Phonon Polaritons in Atomically Thin van der Waals Crystals of Boron Nitride. *Science* **343**, 1125-1129 (2014).
25. Yoxall, E. et al. Direct observation of ultraslow hyperbolic polariton propagation with negative phase velocity. *Nat Photon* **9**, 674-678 (2015).
26. Giles, A.J. et al. Imaging of Anomalous Internal Reflections of Hyperbolic Phonon-Polaritons in Hexagonal Boron Nitride. *Nano Letters* **16**, 3858-3865 (2016).
27. Caldwell, J.D. et al. Sub-diffractional volume-confined polaritons in the natural hyperbolic material hexagonal boron nitride. *Nat Commun* **5** (2014).
28. Xu, X.G. et al. One-dimensional surface phonon polaritons in boron nitride nanotubes. *Nat Commun* **5** (2014).
29. Giles, A.J. et al. Ultralow-loss polaritons in isotopically pure boron nitride. *Nature Materials* **17**, 134 (2017).
30. Li, P. et al. Infrared hyperbolic metasurface based on nanostructured van der Waals materials. *Science* **359**, 892-896 (2018).
31. Ambrosio, A. et al. Selective excitation and imaging of ultraslow phonon polaritons in thin hexagonal boron nitride crystals. *Light: Science & Applications* **7**, 27 (2018).
32. Poddubny, A., Iorsh, I., Belov, P. & Kivshar, Y. Hyperbolic metamaterials. *Nat Photon* **7**, 948-957 (2013).
33. High, A.A. et al. Visible-frequency hyperbolic metasurface. *Nature* **522**, 192-196 (2015).
34. Dai, S. et al. Subdiffractional focusing and guiding of polaritonic rays in a natural hyperbolic material. *Nat Commun* **6** (2015).
35. Li, P. et al. Hyperbolic phonon-polaritons in boron nitride for near-field optical imaging and focusing. *Nat Commun* **6** (2015).
36. Krishnamoorthy, H.N.S., Jacob, Z., Narimanov, E., Kretzschmar, I. & Menon, V.M. Topological Transitions in Metamaterials. *Science* **336**, 205-209 (2012).
37. Dai, S. et al. Graphene on hexagonal boron nitride as a tunable hyperbolic metamaterial. *Nat Nano* **10**, 682-686 (2015).
38. Li, P. et al. Reversible optical switching of highly confined phonon–polaritons with an ultrathin phase-change material. *Nature Materials* **15**, 870 (2016).
39. Chen, J. et al. Optical nano-imaging of gate-tunable graphene plasmons. *Nature* **487**, 77-81 (2012).
40. Fei, Z. et al. Gate-tuning of graphene plasmons revealed by infrared nano-imaging. *Nature* **487**, 82-85 (2012).
41. Liu, M.K. et al. Anisotropic Electronic State via Spontaneous Phase Separation in Strained Vanadium Dioxide Films. *Physical Review Letters* **111**, 096602 (2013).
42. Qazilbash, M.M. et al. Mott Transition in $VO_2$ Revealed by Infrared Spectroscopy and Nano-Imaging. *Science* **318**, 1750-1753 (2007).
43. Liu, M. et al. Symmetry breaking and geometric confinement in VO2: Results from a three-dimensional infrared nano-imaging. *Applied Physics Letters* **104**, 121905 (2014).

44. Dai, S. et al. Efficiency of Launching Highly Confined Polaritons by Infrared Light Incident on a Hyperbolic Material. *Nano Letters* (2017).
45. Alonso-González, P. et al. Controlling graphene plasmons with resonant metal antennas and spatial conductivity patterns. *Science* **344**, 1369-1373 (2014).
46. Fei, Z. et al. Infrared Nanoscopy of Dirac Plasmons at the Graphene–SiO2 Interface. *Nano Letters* **11**, 4701-4705 (2011).
47. Qazilbash, M.M. et al. Infrared spectroscopy and nano-imaging of the insulator-to-metal transition in vanadium dioxide. *Physical Review B* **79**, 075107 (2009).
48. Sternbach, A.J. et al. Artifact free time resolved near-field spectroscopy. *Opt. Express* **25**, 28589-28611 (2017).
49. Koenderink, A.F., Alù, A. & Polman, A. Nanophotonics: Shrinking light-based technology. *Science* **348**, 516-521 (2015).
50. Vakil, A. & Engheta, N. Transformation Optics Using Graphene. *Science* **332**, 1291-1294 (2011).
51. Di Pietro, P. et al. Observation of Dirac plasmons in a topological insulator. *Nature Nanotechnology* **8**, 556 (2013).
52. Nemilentsau, A., Low, T. & Hanson, G. Anisotropic 2D Materials for Tunable Hyperbolic Plasmonics. *Physical Review Letters* **116**, 066804 (2016).
53. Folland, T.G. et al. Reconfigurable Mid-Infrared Hyperbolic Metasurfaces using Phase-Change Materials. *ArXiv e-prints* (2018).

**Acknowledgments:**

Work at UCSD and Columbia University on optical phenomena in vdW materials is supported by DOE-BES DE-FG02-00ER45799 and Betty Moore Foundation's EPiQS Initiative through Grant GBMF4533. Research at UCSD of nano-IR instrumentation is supported by AFOSR FA9550-15-1-0478, and ONR N00014-15-1-2671. Q.M. and P.J.H. were supported by the Center for Excitonics, an Energy Frontier Research Center funded by the DOE, Office of Science, BES under Award Number DESC0001088 and AFOSR grant FA9550-16-1-0382, as well as the Gordon and Betty Moore Foundation's EPiQS Initiative through Grant GBMF4541 to P.J.H.

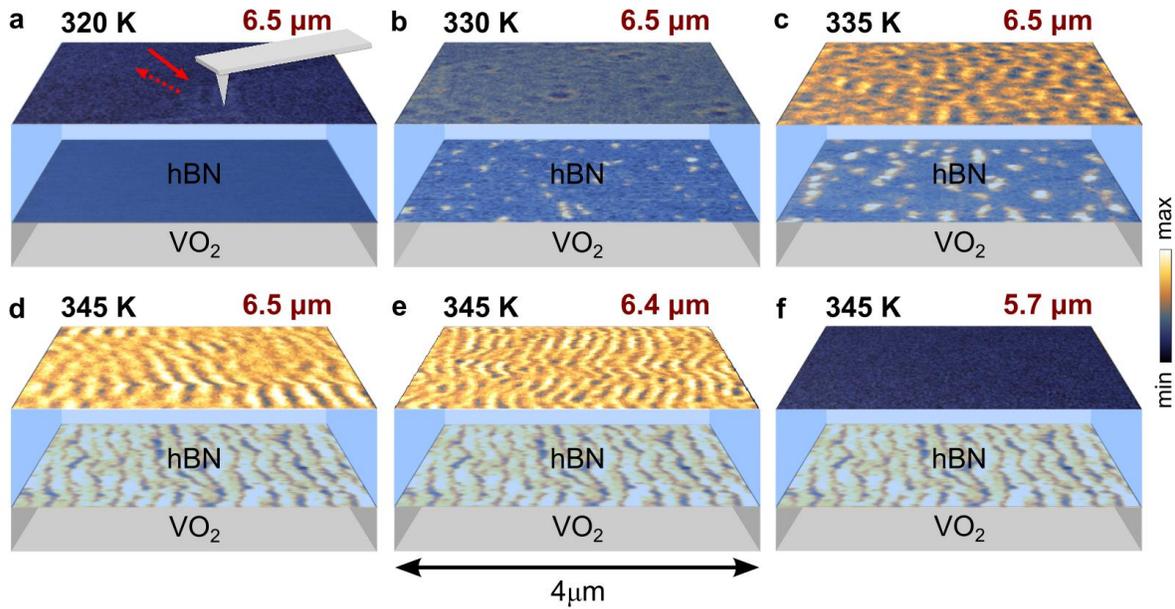

**Figure 1. s-SNOM nano-images reveal the hyperbolic phonon polaritons and insulator to metal phase transition in hBN/VO$_2$ heterostructures.** s-SNOM images of hyperbolic phonon polaritons (plotted on hBN top surface) affected by insulator to metal transition (hBN/VO$_2$ interface) recorded at the temperature and incident IR wavelength of 320K and 6.5 μm (**a**), 330K and 6.5 μm (**b**), 335K and 6.5 μm (**c**), 345K and 6.5 μm (**d**), 320K and 6.4 μm (**e**) and 320K and 6.7 μm (**f**). Thickness of the hBN: 73 nm.

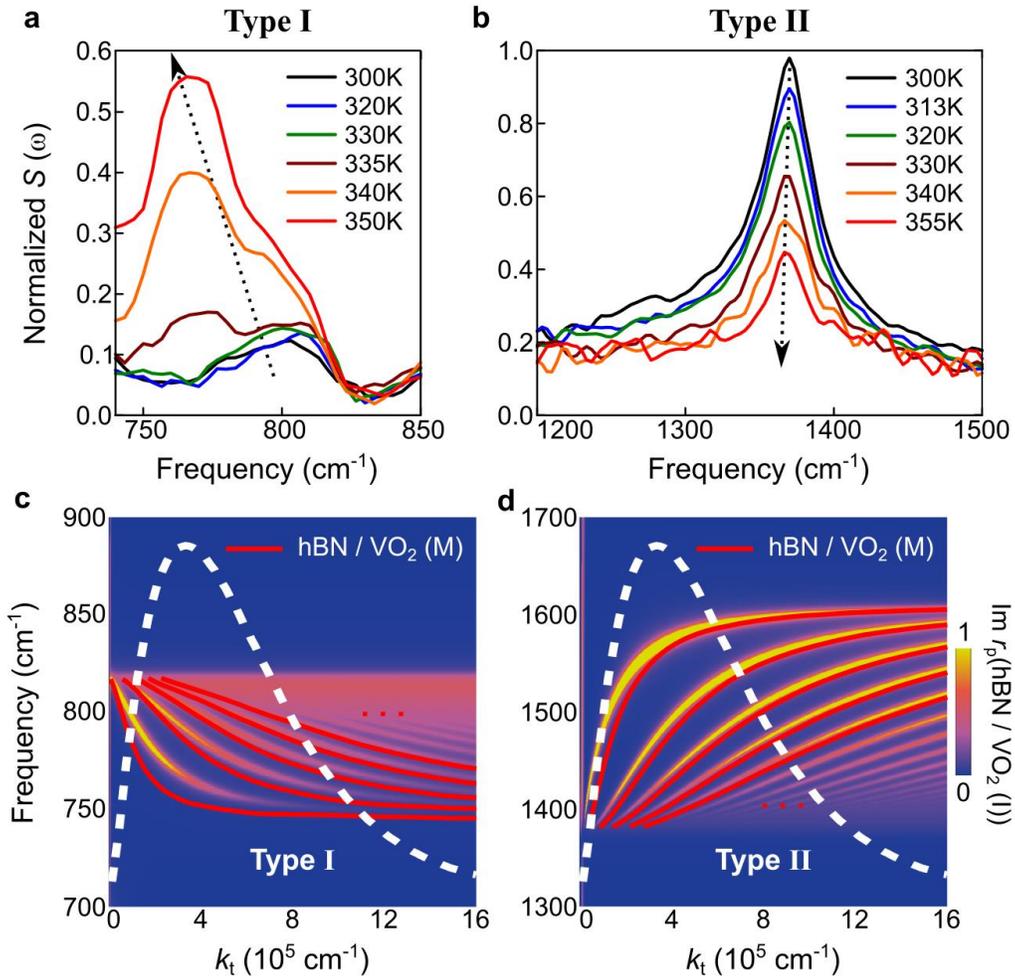

**Figure 2**. **Tuning hyperbolic phonon polaritons in hBN/VO$_2$ heterostructures by temperature control. a**, **b**, nano-FTIR spectra of hBN/VO$_2$ at various temperature in Type I (**a**) and Type II (**b**) hyperbolic region. **c**, **d**, Simulation of the hybridization between hyperbolic phonon polaritons and insulator to metal transition in hBN/VO$_2$ heterostructures in Type I (**c**) and Type II (**d**) hyperbolic region. The polariton dispersion are plotted with red curves at metallic state and with false color at insulating state of VO$_2$, respectively. The white dashed curves indicate the near-field coupling weight function. hBN thickness: 73 nm.